\documentclass[aps,prb,preprint,groupedaddress,showpacs]{revtex4}
\usepackage{amsmath}
\usepackage{graphicx}
\usepackage{dcolumn}
\usepackage{bm}

\begin{document}
\title{A gauge invariant derivation of the AC Josephson frequency and a reconsideration of the origin of the phase of the order-parameter}
\author{Hiroyasu Koizumi}
\email[ ]{koizumi@ims.tsukuba.ac.jp}
\affiliation{Institute of Materials Science, University of Tsukuba,Tsukuba, Ibaraki 305-8573, Japan}
\date{\today}
\begin{abstract}
We derive the AC Josephson frequency using the gauge invariant equations of motion with including the battery-contact effect. The frequency is given as a sum of a contribution from an Aharonov-Bohm phase that arises when charged-partices pass through an electric field in the insulator and that from a chemical potential difference arising form the battery contact; each of them contributes $|q|V/h$ to the frequency where $q$ is the charge, thus, the sum is $2|q|V/h$. The observed Josepshon frequency, $2|e|V/h$, hence, means that the charge on the tunneling particles is $e$. 
A variety of derivations so far miss one of the above two contributions; the original derivation misses the first contribution due to the lack of inclusion of the electric field in the insulator.
The present result indicates that the phase of the order-parameter does not arise from the number fluctuations of Cooper pairs. We present an alternative origin for it; we argue that a Berry phase that arises from spin-vortices is the origin.
\end{abstract}

\pacs{74.50.+r,03.65.Vf,74.20.-z,74.72.-h}
\maketitle
\section{Introduction}
The origin of persistent current in superconductors is believed to be a solved problem \cite{Anderson87}. One of the foundations of the current understanding of the persistent current generation is the appearance of an AC current of a frequency $2|e|V/h$ in a superconductor-insulator-superconductor junction with an application of a constant voltage $V$, where $e$ is the electron charge and $h$ Planck's constant. The prediction of this effect (the AC Josephson effect \cite{Josephson}) and its observation \cite{Shapiro} established the validity of the current understanding. Especially, $2|e|$ in the Josephson frequency $2|e|V/h$ has been considered as the direct evidence that paired-electrons tunnel through the junction in accordance with the BCS theory \cite{BCS}. In this work, however, we show that the frequency $2|e|V/h$ actually indicates that the persistent current carriers are not paired-electrons but electrons.

A variety of derivations so far have defects. A common defect originates from the usage of the
Cohen-Falicov-Philips tunneling Hamiltonian\cite{CFP} without including the electromagnetic field in the insulator. \cite{Josephson69} In the Josephson tunneling case, the voltage-drop occurs across the insulator between the two superconductors of the junction, and an electric field that balances the battery voltage exists in the insulator; when a charged particle passes through the insulator, it is either accelerated or decelerated by the electric field. This situation is markedly different from the ordinary tunneling case where the insulator acts simply as a potential barrier.

The preset result calls for rethinking of the origin of the persistent current generation in superconductors; it indicates that Cooper pairs play only a secondary role as to the persistent current generation is concerned. Since the bulk properties of superconductors are well-explained by the Cooper pair formation, an explanation that is compatible with the above observation is that tunneling particles are broken-Cooper-pairs or unpaired electrons that exist in the surface region. In this respect, it is worth noting that serious deviations from the BCS theory have been observed in the Knight shift measurements; and such deviations have been thought to reflect the difference of electronic states in the bulk and in the surface region\cite{Reif,Knight}. 
  
As an alternative for the conventional persistent current generation mechanism, we will put forward a novel one that has been proposed, recently \cite{koizumi,koizumi2}. It utilizes a Berry phase arising from spin-vortices \cite{Berry}; the persistent current is realized as a coherent collection of stable loop currents that are generated around spin-vortices by the fictitious magnetic field of the Berry phase. \cite{koizumi2}  In the latter half of the present work, we will show that this mechanism (we will call it the ``spin-vortex superconductivity'') explains the observed AC Josephson frequency $2|e|V/h$ \cite{Shapiro}, the flux quantization in the unit $hc/2|e|$  ($c$ is the speed of light) \cite{Deaver}, and the oscillation pattern of the SQUID \cite{Mercereau}. 

If the spin-vortex superconductivity is the right one, the origin of the phase of the order parameter is the Berry phase arising from spin-vortices. In this respect, it is worth noting that there are a number of indications that spin-vortices are present in cuprate superconductors \cite{koizumi2}; thus, the present view may provide a new perspective that leads to the elucidation of the mechanism of the cuprate superconductivity.

\section{A gauge invariant derivation of the AC Josephson Frequency}
Let us derive the AC Josephson frequency in a gauge invariant manner using equations of motion for the order parameter.

First,  we denote the superconducting order parameter as
\begin{eqnarray}
\Psi=\rho^{1/2}e^{i \theta},
\label{orderP}
\end{eqnarray}
where $\rho$ is the supercurrent carrier density and $\theta$ is the phase of the order parameter. In the conventional theory $\rho$ is the number density of Cooper pairs, and $\theta$ is the phase variable that is canonical conjugate to $\rho$.

\begin{figure}
\begin{center}
\includegraphics[scale=0.6]{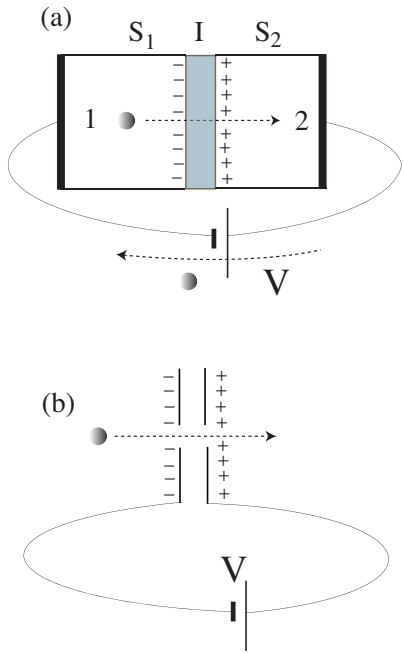}
\end{center}
\caption{(a) A charged particle with charge $q$ tunneling through an superconductor-insulator-superconductor junction that is connected to a battery of the voltage $V$. The junction is a capacitor in the zeroth order approximation, thus, surface charges develop so that the resulting electric field balance the voltage $V$. The particle tunneling from S$_1$ to S$_2$ is accompanied by another charge-transfer from S$_2$ to S$_1$ through the battery which fills the vacant level created by the tunneling electron. (b) A charged particle with charge $q$ passes through the capacitor of the voltage $V$; it acquires the phase shift corresponds to the energy gain of $ |q|V$.}
\label{SIS}
\end{figure}

We consider an SIS (superconductor-insulator-superconductor) junction which is connected to a battery of the voltage $V$ as seen in Fig.~\ref{SIS}(a). In an ideal situation, the voltage-drop that balances the battery voltage occurs entirely across the insulator between the two superconductors, and the resistance is zero. This means that the junction is a capacitor in the zeroth order approximation, thus, surface charges develop so that the generated electric field balances the battery voltage.

The Josephson current is given by
\begin{eqnarray}
J=J_0 \sin \phi,
\label{current}
\end{eqnarray}
where $J_0$ is a constant, and $\phi$ is expressed as
\begin{eqnarray}
\phi= \int_1^2 \left( {q \over {\hbar c}}  {\bf A} -\nabla \theta \right)\! \cdot \!d{\bf r};
\label{phi1}
\end{eqnarray}
$\hbar$ is Planck's constant divided by $2\pi$, $q$ is the charge on the current carrier, and ${\bf A}$ is the vector potential; the integration is performed along a path that connects points in the superconductors S$_1$ and S$_2$ through the insulator I.

The combination
\begin{eqnarray}
\nabla \theta - {q \over {\hbar c}}  {\bf A}
\label{gaugeA}
\end{eqnarray}
is gauge invariant since for the gauge transformation
\begin{eqnarray}
{\bf A}'={\bf A} + \nabla f,
\label{gauget}
\end{eqnarray}
$\theta$ transforms as
\begin{eqnarray}
\theta'=\theta +{q \over {\hbar c}}f,
\label{gauge}
\end{eqnarray}
where $f$ is a single-valued function.
The current in Eq.~(\ref{current}) is gauge invariant, accordingly.

The following combination is also gauge invariant
\begin{eqnarray}
\dot{\theta} + {{q\varphi}  \over {\hbar}} 
\label{inv}
\end{eqnarray}
because the scalar potential $\varphi$ transforms as 
$\varphi'$=$\varphi -\dot{f}/c$. 
In the following, the gauge invariant combination
in Eq.~(\ref{inv}) plays a crucial role.

The basic equations of superfluid phenomena \cite{Anderson66} are given by
\begin{eqnarray}
\dot{\rho}&=&{ {1} \over {\hbar}} {  {\delta {\cal H}} \over {\delta \theta}}, 
\label{boseA}
\\
\dot{\theta}&=&-{1 \over \hbar} {{\delta {\cal H}} \over {\delta \rho}}
\label{bose}
\end{eqnarray}
where ${\cal H}$ is an energy functional.

For the superconductivity case, the above equations should be modified as
\begin{eqnarray}
\dot{\rho}&=&{ {1} \over {\hbar}} {  {\delta \bar{\cal H}} \over {\delta \theta}},
\label{EOM0}
\\
\dot{\theta}+{q \over \hbar} \varphi &=&-{1 \over \hbar} {{\delta \bar{\cal H}} \over {\delta \rho}},
\label{EOM1}
\end{eqnarray}
to take into account the interaction of electromagnetic field and charged-particles. Here, 
$ \bar{\cal H}$ is a gauge invariant energy functional,
$\bar{\cal H}=\bar{\cal H}[\rho, \nabla \theta \!-\! {q \over {\hbar c}}{\bf A}]$, and ${{\delta \bar{\cal H}} \over {\delta \rho}}$ is a gauge invariant chemical potential, in which $\nabla \theta$ appears in the gauge invariant combination $\nabla \theta\!-\! {q \over {\hbar c}}{\bf A}$.

Let us calculate the time-derivative of $\phi$. Using Eqs.~(\ref{phi1}) and (\ref{EOM1}), and the relation
${\bf E}$=$ -\partial_t {\bf A}/c\!-\! \nabla \varphi$,
we obtain
\begin{eqnarray}
\dot{\phi}= -{q \over \hbar} \int_1^2 {\bf E} \cdot d{\bf r}+{ 1 \over \hbar} \left( \left.  {{\delta \bar{\cal H}} \over {\delta \rho}}\right|_2- \left. {{\delta \bar{\cal H}} \over {\delta \rho}}\right|_1 \right).
\label{phit}
\end{eqnarray}

There are two contributions to it. The first is a contribution that arises when a charged-particle passes through the electric field in the insulator (Fig.~\ref{SIS}(b)). Since the electric field by the surface charges balances the voltage of the battery, it is calculated as
\begin{eqnarray}
-{q \over \hbar} \int_1^2 {\bf E} \cdot d{\bf r}={{qV} \over \hbar }.
\label{con1}
\end{eqnarray}
This is the Aharonov-Bohm phase contribution. \cite{Aharonov}
This term is absent in the original derivation; \cite{Josephson,Josephson69}
there, the insulator is taken as a potential barrier just like an ordinary tunneling problem. \cite{CFP}

The second is a contribution from the gauge invariant chemical potential difference produced by the battery contact. Since the battery maintains the chemical potential difference $qV$, it is calculated as
\begin{eqnarray}
{ 1 \over \hbar} \left( \left.  {{\delta \bar{\cal H}} \over {\delta \rho}}\right|_2- \left. {{\delta \bar{\cal H}} \over {\delta \rho}}\right|_1 \right)={{qV} \over \hbar }.
\label{mudiff}
\end{eqnarray}
The constant voltage is maintained through the charge flow into and out of the junction (see Fig.~\ref{SIS}(a)); thereby, the work done by the battery is added to the electronic system in the junction. This contribution will be missed if the junction is treated as an isolated system. 

Overall, the time-derivative of $\phi$ is given by $\dot{\phi}={{2qV} / \hbar }$,
which corresponds to the Josephson frequency 
\begin{eqnarray}
f_{\rm J}={{2|q|V} \over h }.
\label{phit2}
\end{eqnarray}
The observed frequency\cite{Shapiro} is $2|e|V/h$, thus, $q$ should be $e$. This means that the charge on the tunneling particle is $e$, not $2e$. This reverses the current understanding that the Cooper-pair tunneling is responsible for $f_{\rm J}$.

\begin{figure}
\begin{center}
\includegraphics[scale=0.6]{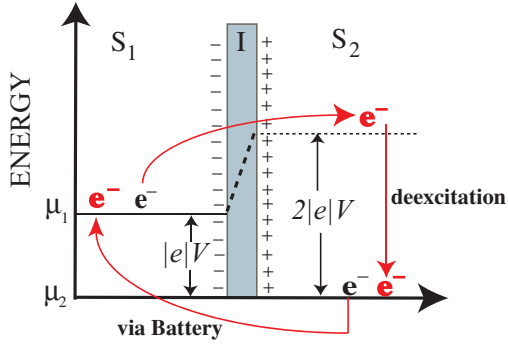}
\end{center}
\caption{Energy diagram for an electron tunneling through the SIS junction in Fig.~\ref{SIS}. $\mu_1$ and $\mu_2$ are chemical potentials in $S_1$ and $S_2$, respectively. One of the most simple tunneling cases is depicted; initially, the tunneling-electron has the energy $\mu_1$. It gains the energy $|e|V$ from the acceleration by the electric field in I. The electron tunneling is accompanied by an electron transfer from S$_2$ to S$_1$ via the battery; this gives rise to the energy gain $\mu_1-\mu_2$. The total energy gain for the electronic system by this process is, thus, $|e|V+\mu_1-\mu_2=2|e|V$. When this energy is discarded, the transfered electron fills the hole in S$_2$ created by the electron transfer through the battery. Black and red electrons schematically describe electrons before and after the process, respectively.}
\label{SISenergy}
\end{figure}

In order to clarify the above rather abstract argument, let us examine the same problem from the energy point of view by identifying the charged particle as an electron. In Fig.~\ref{SISenergy}, the energetics for the tunneling process is depicted. The chemical potentials in S$_1$ and S$_2$ are  $\mu_1$ and $\mu_2$, respectively. When the tunneling process is absent, the battery creates the situation where all states below $\mu_1$ and $\mu_2$ are occupied in S$_1$ and in S$_2$, respectively. 

When the tunneling from S$_1$ to S$_2$ through the insulator occurs, following events happen, roughly: First, the tunneling-electron is accelerated by the electric field inside the insulator, and gains the energy $|e|V$ (this energy is taken from the energy stored in the electromagnetic filed in the insulator). Second, the hole in S$_1$ created by the tunneling is filled by an electron that is transferred from S$_2$ to S$_1$ through the battery; the energy gain by this process is $\mu_1- \mu_2=|e|V$, which is the work done by the battery. The total energy gain is, thus, $|e|V+\mu_1- \mu_2=2|e|V$. Third, by deexcitation of the energy $2|e|V$, the tunneled-electron fills a hole in S$_2$ that is created by the via-battery-electron-transfer. If the discarded energy is used for an excitation of an electron in S$_2$, a reverse tunneling process will start. 
Actually, there are many ways to fill the holes created by the tunneling and the via-battery-electron-transfer; thus, the real processes are much more complicated; nevertheless, the net energy gain or loss is $2|e|V$ for each tunneling process.
  
\section{Models for the Josephson junction}
The result obtained in the previous section is surprising; it reverses the current understanding that the Cooper-pair-tunneling is the origin of the observed AC Josephson  frequency. 
Here, we consider it, again, using a textbook model \cite{Feynman}. This will make the general argument given above clearer.

The system of equations of the model is given by
\begin{eqnarray}
i{\hbar} {{\partial } \over {\partial t}} \Psi_1&=&T\Psi_2+\mu_1\Psi_1
\label{Kittel1A}
\\
i{\hbar} {{\partial } \over {\partial t}} \Psi_2&=&T\Psi_1+\mu_2\Psi_2,
\label{Kittel1}
\end{eqnarray}
where $\Psi_1\!=\!\rho_1^{1/2} e^{i \theta_1}$ and $\Psi_2\!=\!\rho_2^{1/2} e^{i \theta_2}$ 
are order-parameters in S$_1$ and S$_2$, respectively; $T$ is a transfer integral between the two superconductors through the insulator; and
$\mu_1$ and $\mu_2$ are chemical potentials of S$_1$ and S$_2$, respectively. The battery produces the chemical potential difference $\mu_2-\mu_1\!=\!qV$; this gives rise to the contribution that corresponds to Eq.~(\ref{mudiff}).

When a battery is connected to the junction, the equations in Eqs.~(\ref{Kittel1A}) and (\ref{Kittel1}) have to be modified; in order to include the electric field generated by the surface charges, scalar and vector potentials must be introduced,
\begin{eqnarray}
i{\hbar} {{\partial } \over {\partial t}} \Psi_1&=&Te^{i {q \over {\hbar c}} \int^2_1 {\bf A} \cdot
d{\bf r}}\Psi_2+q\varphi_1 \Psi_1 +\mu_1\Psi_1
\label{Kittel2A}
\\
i{\hbar} {{\partial } \over {\partial t}} \Psi_2&=&Te^{i {q \over {\hbar c}} \int^1_2 {\bf A} \cdot
d{\bf r}}\Psi_1+q\varphi_2 \Psi_2+\mu_2\Psi_2.
\label{Kittel2}
\end{eqnarray}
Actually, the inclusion of the vector and scalar potentials is crucial to obtain gauge invariant results, as seen below.

It should be remembered that Eqs.~(\ref{Kittel2A}) and (\ref{Kittel2}) do not include the charge flow into and out of the junction; such a process has to be included ``by hand''.

\begin{figure}
\begin{center}
\includegraphics[scale=0.6]{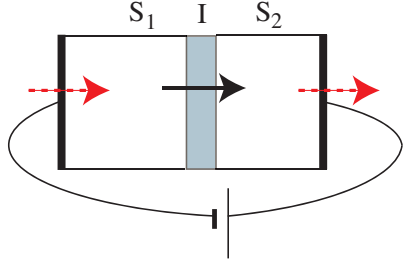}
\end{center}
\caption{The charge flow between the two superconductors (solid-line arrow), and  those added ``by hand'' (dotted-line arrows) to take into account the battery contact effect. }
\label{SISbyhand}
\end{figure}

From Eqs.~(\ref{Kittel2A}) and (\ref{Kittel2}), equations for $\rho_1,\rho_2, \theta_1$, and $\theta_2$ are obtained;
\begin{eqnarray}
\dot{\rho}_1\!&\!=\!&\!-\!\dot{\rho}_2\!=\!{{2T} \over {\hbar}} (\rho_1 \rho_2)^{1/2}
\sin \left(\!\int^2_1\!( {q \over {\hbar c}}  {\bf A} \! - \! \nabla \theta) \! \cdot \! d{\bf r}\!\right)
\label{Kittel3A}
\\
\dot{\theta}_1 \!+\! {q \over {\hbar}} \varphi_1\!&\!=\!& \!-{{\mu_1} \over {\hbar}} 
\!-\!{T \over \hbar}\left(\!{ {\rho_2} \over {\rho_1}} \right)^{1/2} \!\cos \!\left(\int^2_1\!( {q \over {\hbar c}}  {\bf A} \! - \! \nabla \theta) \!\cdot \!d{\bf r}\!\right)
\label{Kittel3B}
\\
\dot{\theta}_2 \!+\! {q \over {\hbar}} \varphi_2\!&=&\!-{{\mu_2} \over {\hbar}} 
\!-\!{T \over \hbar}\left(\!{ {\rho_1} \over {\rho_2}} \right)^{1/2} \!\cos \! \left(\int^2_1\!( {q \over {\hbar c}}  {\bf A} \! - \! \nabla \theta) \!\cdot \!d{\bf r}\!\right)\!.
\label{Kittel3}
\end{eqnarray}
The r.h.s. of Eq.~(\ref{Kittel3A}) corresponds to Eq.~(\ref{current}); actually, if we take into account the charge flow into and out of the junction ``by hand'' as depicted in Fig.~\ref{SISbyhand}, the condition $\dot{\rho}_1$=$\dot{\rho}_2$=$0$ is established. The second and third correspond to Eq.~(\ref{EOM1}); by imposing the condition $\rho_1=\rho_2$  ``by hand'', the gauge invariant chemical potential difference across the junction is maintained to be $qV$. Then, the relation in Eq.~(\ref{mudiff}) is obtained. Thus, the final result is Eq.~(\ref{phit2}).

It may be worth noting that $\mu$ cannot be replaced by $q\varphi$; $\mu$ is gauge invariant but $q\varphi$ is not. The gauge invariant chemical potential difference brought about by the battery-contact has to be included using $\mu$; this point becomes apparent if we adopt the gauge in which $\varphi=0$ and take the limit $T \rightarrow 0$ in Eqs.~(\ref{Kittel2A}) and (\ref{Kittel2}); then, the only way to include the chemical potential difference is to use $\mu_1$ and $\mu_2$.

\begin{figure}
\begin{center}
\includegraphics[scale=0.8]{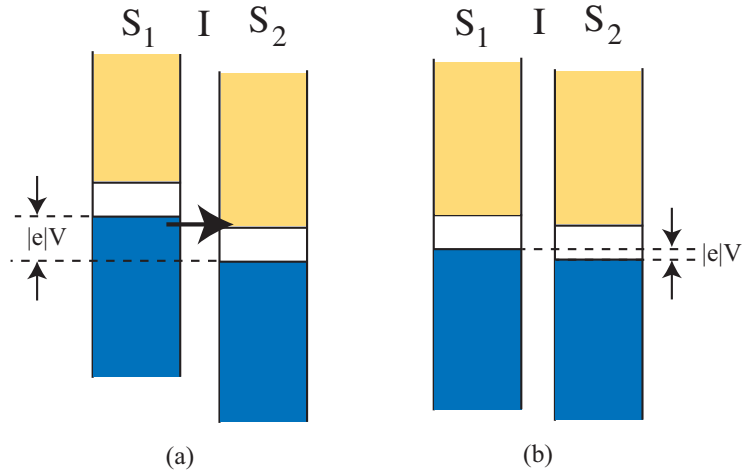}
\end{center}
\caption{Schematic plots of density of states. (a) the SIS junction biased by a voltage V that allows tunneling from the filled band in S$_1$ to the empty band in S$_2$. The filled and empty bands are colored in blue and yellow, respectively. The DC current flow with dissipation occurs. The junction is like a resistor in this case; (b) the same as in (a) but the voltage is not enough to allow tunneling from the filled band in S$_1$ to the empty band in S$_2$. The voltage of the battery is supported by the surface charges developed in the interfaces. The junction is like a capacitor in this case.}
\label{Tunnel}
\end{figure}

A very often used model for the SIS junction is the Cohen-Falicov-Phillips tunneling Hamiltonian \cite{CFP} given by
\begin{eqnarray}
H_{\rm CFP}=H_{\rm L}+ H_{\rm R}+ H_{\rm T}
\end{eqnarray}
where $H_{\rm L}$,  $H_{\rm R}$, and $H_{\rm T}$ are the Hamiltonians for the left superconductor S$_1$, right superconductor S$_2$, and the tunneling through the insulator, respectively.

Two types of tunneling currents in the SIS junction are currently explained the above Hamiltonian.
One is the single-particle (or quasiparticle) tunneling current, and the other is the current generated by the collective mode $\theta$ (usually, it is called, `` the Cooper-pair tunneling", but we will not use it, since it is reversed in the preset work). Both of them are usually calculated by taking $H_{\rm T}$ as a perturbation.

The single particle current is dominant in the situation given in Fig.~\ref{Tunnel}(a).
This yields the current that reflects the density of states of the two superconductors including the chemical potential difference brought about by the battery-contanct;\cite{Ketterson}
\begin{eqnarray}
J={{ 4 \pi e } \over {\hbar}} \sum_{l, r} |T_{lr}|^2 \left( f_l -f_r \right) \delta(E_l-eV-E_r)
\end{eqnarray}
where $l$ and $r$ denotes the quasiparticle states in S$_1$ and S$_2$, respectively; their energies are $E_l$ and $E_r$, respectively. $T_{lr}$ are tunneling matrix elements, and $f_l$ and $f_r$ are distribution functions for the particles in the states $l$ and $r$, respectively.

Here, the effect that corresponds to Eq.~(\ref{mudiff}) is taken into account by the energy shift $eV$ in the argument of the delta function. In this case, a DC current flows, and the junction is more like a resistor than a capacitor. The phase $\theta$ is not definable throughout the junction. The matrix element $T_{lr}$ simply describes the potential barrier penetration rate through the insulator; thus, the phase effect that corresponds to Eq.~(\ref{con1}) does not arise. 

On the other hand, the other type of current will be dominant in the situation given in Fig.~\ref{Tunnel}(b), in which the phase $\theta$ is definable throughout the junction; this corresponds to the Josephson effect case. The junction is like a capacitor and an AC current flows (the DC Josephson case should be viewed as the $V\rightarrow 0$ limit of the AC case). In the conventional theory, $\theta$ arises as the angle-variable that is canonical conjugate to the Cooper-pair number density. The angle-variable is included in the BCS wave function as
\begin{eqnarray}
|\Psi_{\rm BCS} \rangle=\prod_{ {\bf k}}\left( u_{{\bf k}}+e^{i\theta}v_{{\bf k}}c^{\dagger}_{{\bf k} \uparrow}c^{\dagger}_{ -{\bf k} \downarrow} \right)|{\rm vac}\rangle,
\label{BCS0}
\end{eqnarray}
where the parameters satisfy $u^2_{ {\bf k}}+v^2_{ {\bf k}}=1$. $\theta$ has been thought to be canonical conjugate to the number density of Cooper-pairs, thus, the Josephson current is often said to be the Cooper-pair tunneling current.

Temporal and spatial variations of $\theta$ with added vector and scalar potentials give rise to the additional contribution given in Eq.~(\ref{con1}); the magnetic field counterpart caused by the spatial variation of $\theta$ with the added vector potential is known to give rise to a very sensitive current flow to an applied magnetic field as is seen in the SQUID experiment.

In the original derivation for the AC Josephson frequency, the contribution given in Eq.~(\ref{con1}) is missing;\cite{Josephson,Josephson69} this may have occurred by the direct use of the theory that is suitable for the quasiparticle tunneling case to the Josephson tunneling case;\cite{Josephson69} then, the phase factor for the Aharonov-Bohm phase was not included. There are other types of defects in the derivations: sometimes, the substitution of $\mu$ (which is gauge invariant) by $q\phi$ (which is gauge covariant) occurs, tacitly; \cite{Anderson69} in this case, the contribution that corresponds to Eq.~(\ref{mudiff}) disappears. Yet, another mistake is related to treating the junction as an isolated system; then, the contribution that corresponds to Eq.~(\ref{mudiff}) also disappears.\cite{Landau}

\section{Spin-vortex superconductivity}
In the conventional superconductivity theory, the phase of the order parameter, $\theta$, is thought to arise from the fluctuation of the number of Cooper-pairs \cite{Anderson66}. In this interpretation, however, the charge on the tunneling particle is $2e$, thus, contradicts the experimental result according to arguments in previous sections.  In the following, we present an alternative origin of the order parameter. Actually, we will show that an electronic system with spin-vortices has $\theta$ and $\rho$ that satisfy the equations of motion in Eqs.~(\ref{EOM0}) and (\ref{EOM1}) with the $e$-charge carrier.

Let us derive the phase $\theta$ that appears in the system with spin-vortices. 

We consider the Hamiltonian given by
\begin{eqnarray}
{H}\!=\!-\!\sum_{ k,j, \sigma} t_{kj}e^{{{iq} \over {\hbar c}} \int^k_j {\bf A}\cdot d{\bf r}}c_{k\sigma}^{\dagger}c_{j \sigma}\!+\!q\sum_{j, \sigma}\varphi_j c^{\dagger}_{j \sigma}c_{j\sigma}\!+\!H_{\rm int}
\label{Hub}
\end{eqnarray}
where the first term describes electron hopping, the second term is that of the scalar potential, and the third term is the rest. 

When spin-vortices are present the following new electron operators $a_j$ and $b_j$ are convenient;\cite{koizumi} they are related to the original operators, $c_{j \uparrow}$ and $c_{j \downarrow}$ as 
\begin{eqnarray}
\left( \begin{array}{c}
a_{j}\\
b_{j}
\end{array}
\right)={ {e^{ i{\chi_j \over 2}}} \over \sqrt{2} } \left(
\begin{array}{cc}
e^{i {\xi_j \over 2}} & e^{-i {\xi_j \over 2}} \\
-e^{i {\xi_j \over 2}} & e^{-i {\xi_j \over 2}} 
 \end{array}
 \right) \left( \begin{array}{c}
c_{j \uparrow}\\
c_{j \downarrow}
\end{array}
\right).
\label{transform1}
\end{eqnarray}

Spin operators at the $j$th site are given by
\begin{eqnarray}
S_x(j)&=&{1 \over 2} \cos \xi_j (a_j^{\dagger}a_j-b_j^{\dagger}b_j) +{i \over 2} \sin \xi_j (a_j^{\dagger}b_j-b_j^{\dagger}a_j) ,
\\
S_y(j)&=&{1 \over 2} \sin \xi_j (a_j^{\dagger}a_j-b_j^{\dagger}b_j) -{i \over 2} \cos \xi_j (a_j^{\dagger}b_j-b_j^{\dagger}a_j) ,
\\
S_z(j)&=&-{1 \over 2}  (a_j^{\dagger}b_j+b_j^{\dagger}a_j).
\end{eqnarray}

If we express expectation values of $a_j^{\dagger}a_j\!-\!b_j^{\dagger}b_j$ and $i (a_j^{\dagger}b_j\!-\!b_j^{\dagger}a_j)$ as $\langle a_j^{\dagger}a_j\!-\!b_j^{\dagger}b_j \rangle \!=\!A_j$ and
$i\langle a_j^{\dagger}b_j\!-\!b_j^{\dagger}a_j \rangle\!=\!B_j$, respectively, 
we have
\begin{eqnarray}
\langle S_x(j) \rangle&=& \sqrt{A_j^2 + B_j^2} 
\cos ( \xi_j -\alpha_j) ,
\\
\langle S_y(j) \rangle&=& \sqrt{A_j^2 + B_j^2} 
\sin ( \xi_j -\alpha_j) ,
\end{eqnarray}
where $\alpha_j= \tan^{-1} B_j/A_j$. Above equations indicate that
the phase $\xi_j -\alpha_j$ is the spin direction in the $x$-$y$ plane, thus, $\xi$ and $\xi+2\pi$ are physically equivalent. 
The phase factor $\exp( i{\chi_j / 2})$ in Eq.~(\ref{transform1}) is introduced to ensure the single-valuedness of the transformation matrix by compensating the sign-change of $\exp(\pm i {\xi_j / 2})$ that occurs for the shift $\xi \rightarrow \xi + 2\pi$. We may take $\chi=\xi$, but other choices are also possible. 

For the gauge transformation in Eq.~(\ref{gauget}),
electron operators are modified as
\begin{eqnarray}
c_{j \sigma}\rightarrow \exp \!\left( -i { q \over {\hbar c}}f_j \right) c_{j \sigma},
\end{eqnarray}
which, according to Eq.~(\ref{transform1}), means that $\chi$ is modified as
\begin{eqnarray}
\chi'_{j } = \chi_{j }- { {2q} \over {\hbar c}}f_j;
\label{chi}
\end{eqnarray}
therefore we may take $\theta$ in Eq.~(\ref{gauge}) to be $\theta=-\chi/2$.

Using the new creation and annihilation operators, the hopping term in the Hamiltonian Eq.~(\ref{Hub}) is now written as
\begin{eqnarray}
-\! \!\sum_{k,j}\! t_{kj}
e^{ i\int^k_j ({q \over {\hbar c}}{\bf A}\!+\!{{\nabla \chi} \over 2})\cdot 
d{\bf r}} \Big[
\cos{{\xi_k\!-\!\xi_j} \over 2}(a_{k}^{\dagger} a_{j}\!+\!b_{k}^{\dagger} b_{j})
+ i
\sin{{\xi_k\!-\!\xi_j} \over 2}(\!a_{k}^{\dagger} b_{j}\!+\!b_{k}^{\dagger} a_{j}\!)\Big].
\label{hopping}
\end{eqnarray}
Note that ${\bf A}$ and $\nabla \chi$ appear in the gauge invariant combination $(q {\bf A}/ \hbar c+{{\nabla \chi} / 2})$.
 
Now, let us derive equations of motion for the phase $\chi$ and the number density $\rho$ by employing the time-dependent variational principle with the following Lagrangian, 
\begin{eqnarray}
{\cal L}=\langle \Psi| \left(i \hbar { {\partial} \over {\partial t} } - H \right)|\Psi \rangle,
\label{lag1}
\end{eqnarray}
where, the state vector $|\Psi\rangle$ is expressed as
\begin{eqnarray}
|\Psi \rangle=\sum_{\alpha} f_{\alpha}|\alpha\rangle;
\end{eqnarray}
$\alpha=(n_{a1},\cdots, n_{aN_s}; n_{b1}, \cdots, n_{bN_s})$ denotes an occupation pattern of electrons, where $N_s$ is the number of sites, $n_{aj}$ and $n_{bj}$ are eigenvalues of the number operators $a^{\dagger}_j a_j$ and $b^{\dagger}_j b_j$, respectively. The state vector $|\alpha\rangle$
is defined as
\begin{eqnarray}
|\alpha\rangle=\prod_{j \in S^{\alpha}_a} a^{\dagger}_j \prod_{k \in S^{\alpha}_b} b^{\dagger}_k |{\rm vac}\rangle,
\end{eqnarray}
where $S^{\alpha}_a$ and $S^{\alpha}_b$ denote sets of sites where the occupation numbers in $a^{\dagger}_j a_j$ and $b^{\dagger}_j b_j$ are $1$, respectively.

Noting that $a^{\dagger}_j$ and $b^{\dagger}_j$ are time-dependent because $\chi_j$ and $\xi_j$ are time-depedent, the equation (\ref{lag1}) yields, 
\begin{eqnarray}
{\cal L}\!=\! i \hbar \sum_{\alpha} f_{\alpha}^{\ast}\dot{f}_{\alpha}\!+\!\sum_j {\hbar \over 2}\dot{\chi_j}\rho_j
\!-\!\sum_{\alpha,\beta} f_{\alpha}^{\ast}f_{\beta} \langle \alpha|H|\beta \rangle,
\label{lag2}
\end{eqnarray}
where terms with the time derivative of $\xi$ cancel out.

From the above Lagrangian, the Hamiltonian is obtained as
\begin{eqnarray}
{\cal H}\!=\! \sum_{\alpha,\beta} f_{\alpha}^{\ast}f_{\beta} \langle \alpha|H|\beta \rangle.
\label{ham2}
\end{eqnarray}

Now, consider the ground state for given $\nabla \xi$, $\nabla \chi$, ${\bf A}$, and $\varphi$. The Hohenberg-Kohn theorem \cite{hohenberg} tells us that it is a functional of $\rho$; then, we may express it as
\begin{eqnarray}
\sum_{\alpha,\beta} f_{\alpha}^{\ast}f_{\beta} \langle \alpha|H|\beta \rangle
\!=\!\bar{\cal H}[\rho, \nabla \chi \!+\! {{2q} \over {\hbar c}}{\bf A}, \nabla \xi]\!+\!\sum_{j}  q \varphi_j \rho_j
\label{ham3}
\end{eqnarray}
where the Coulomb potential term is separately treated for the later convenience.

Next, we make a crucial assumption: we consider the situation where low energy excitations that give rise to the time dependence of $f_{\alpha}$ are absent; this means that the interaction Hamiltonian $H_{\rm int}$ is such that it gives rise to an energy gap that suppresses the time dependence of $f_{\alpha}$. Actually, this is indeed what is observed in the model considered in our previous work \cite{koizumi}. The energy gap formation by the BCS theory may be considered as such a kind.

Then, low energy excitations are described by the following Lagrangian: 
\begin{eqnarray}
{\cal L}\!=\! \sum_j {\hbar \over 2}\dot{\chi_j}\rho_j
\!-\! \bar{\cal{H}}\left [\rho,{\bf A}\!+\!{ {\hbar c} \over {2q}} \nabla \chi, \nabla \xi \right]\!-\!\sum_{j}  q \varphi_j \rho_j.
\label{lag3}
\end{eqnarray}

The momentum conjugate to $\chi_j$ is, thus, obtained as
\begin{eqnarray}
p_{\chi_j}={{\delta {\cal L}} \over {\delta \dot{\chi}_j}}={{\partial {\cal L}} \over {\partial \dot{\chi}_j}}={\hbar \over 2}\rho_j;
\label{pchi}
\end{eqnarray}
this shows that $\rho$ and $-\chi / 2 $ are conjugate variables.

Finally, the equations of motion are given by
\begin{eqnarray}
\dot{\rho}&=&-{ {2} \over {\hbar}} {  {\delta \bar{\cal H}} \over {\delta \chi}}
\\
\dot{\chi}-{{2q} \over \hbar} \varphi &=&{2 \over \hbar} {{\delta \bar{\cal H}} \over {\delta \rho}}
\label{EOM}, 
\end{eqnarray}
which correspond to Eqs.~(\ref{EOM0}) and (\ref{EOM1}) with $\theta=-\chi / 2$.

\section{Flux quantization in the spin-vortex superconductivity}
It has been thought that the quantization of the flux in the unit 
$\Phi_0=hc/2|e|$ \cite{Deaver} is the evidence that the persistent current carriers are Cooper pairs. However, we will show that the same flux quantization is explained by the spin-vortex superconductivity.

In the spin-vortex superconductivity, only 
$\nabla \chi+ {{2q} \over {\hbar c}}{\bf A}$ and $\rho$ are active variables in low energy excitation. In the ground state, it is expected that $\nabla \chi+ {{2q} \over {\hbar c}}{\bf A}$ is zero and  the electric current is zero. If the Meissner effect occurs, the electric current is absent deep inside the sample; then, the condition $\nabla \chi+ {{2q} \over {\hbar c}}{\bf A}=0$ should be established deep inside the sample. 

Let us consider a ring-shaped sample and consider the line-integration of ${\bf A}$ along a path that is deep inside the sample. If this path is so chosen that it encircles the hole of the ring, we have
\begin{eqnarray}
\oint {\bf A} \cdot d{\bf r}={{\hbar c} \over {2|e|}} \oint \nabla \chi \cdot d{\bf r}={{h c} \over {2|e|}}n
\end{eqnarray}
since $\chi$ is an angular variable with period $2\pi$; here, $q=-|e|$ is substituted. The above formula shows that the flux is quantized in the unit $\Phi_0=hc/2|e|$ as is observed in experiments. \cite{Deaver}

The assumed Meissner effect above will arise from the stability  against the fluctuation of $\nabla \chi+ {{2q} \over {\hbar c}}{\bf A}$; a nonzero $\nabla \chi+ {{2q} \over {\hbar c}}{\bf A}$ should increase the total energy, thus, expelling of the magnetic field from the bulk by keeping $\nabla \chi\!+\! {{2q} \over {\hbar c}}{\bf A}\!=\!0$ is energetically favorable.  This expelling of the magnetic field is nothing but the Meissner effect.

\section{AC Josephson effect in the spin-vortex superconductivity}
Now we consider the AC Josephson effect in the spin-vortex superconductivity.

By introducing the ``order parameter'' $\Psi_s$ defined as
\begin{eqnarray}
\Psi_s=\rho^{1/2}e^{-i\chi/2},
\end{eqnarray}
the Lagrangain in Eq.~(\ref{lag3}) is expressed as
\begin{eqnarray}
{\cal L}=\int d^3r \Psi_s^{\ast} \left( i \hbar {{\partial} \over {\partial t}}-\hat{h}-q\varphi\right) \Psi_s
\label{hhat}
\end{eqnarray}
where 
$\hat {h}$ is defined as the operator that satisfies
\begin{eqnarray}
\int d^3r \Psi_s^{\ast} \hat{h} \Psi_s=\bar{\cal{H}}\left [\rho,{\bf A}\!+\!{ {\hbar c} \over {2q}} \nabla \chi, \nabla \xi \right],
\end{eqnarray}
and the conservation of the number of particles 
$\int d^3r \dot{\rho}=0$ is used.

From the stationary condition of the Lagrangian in Eq.~(\ref{hhat}) with respect to the variation of $\Psi_s^{\ast}$, we obtain a ``Schr\"{o}dinger equation'',
\begin{eqnarray}
i \hbar {{\partial} \over {\partial t}}\Psi_s=\left(\hat{h}+q\varphi\right) \Psi_s
\end{eqnarray}
where $\Psi_s$ is the ``wave function'', and $(\hat{h}+q\varphi)$ is the ``Hamiltonian''.

For the Josephson junction problem, we introduce {\it supports} $|1\rangle$ and $|2 \rangle$ in the superconductors $S_1$ and  $S_2$, respectively; they satisfy
\begin{eqnarray}
1&=&\langle 1 | 1 \rangle =  \langle 2 | 2 \rangle,
\nonumber
\\
0&=&\langle 1 | 2 \rangle =  \langle 2 | 1 \rangle, 
\end{eqnarray}
and $\Psi_s$ is expressed as a sum of two functions in the two superconductors;
\begin{eqnarray}
\Psi_s= \Psi_{s1}| 1 \rangle + \Psi_{s2}| 2 \rangle 
\end{eqnarray}
where $\Psi_{s1}\!=\!\rho_1^{1/2} e^{-i \chi_1}$ and $\Psi_{s2}\!=\!\rho_2^{1/2} e^{-i \chi_2}$ are ``wave functions'' for the two superconductors in the junction.

By expressing the ``Hamiltonian '' as 
\begin{eqnarray}
\hat{h}\!+\!q\varphi
\!=\!Te^{ {{iq} \over {\hbar c}} \int^2_1 {\bf A} \cdot
d{\bf r}} |1 \rangle \langle 2 |
\!+\!Te^{{{iq} \over {\hbar c}} \int^1_2 {\bf A} \cdot
d{\bf r}} |2 \rangle \langle 1 |
\!+\!( q\varphi_1 \!+\! \mu_1) |1 \rangle\langle 1 |
\!+\!( q\varphi_2 \!+\! \mu_2) |2 \rangle\langle 2 |,
\end{eqnarray}
we obtain the model Schr\"{o}dinger equation given in Eqs.~(\ref{Kittel2A}) and (\ref{Kittel2});
\begin{eqnarray}
i{\hbar} {{\partial } \over {\partial t}} \Psi_{s1}\!&=&\!Te^{i {q \over {\hbar c}} \int^2_1 {\bf A} \cdot
d{\bf r}}\Psi_{s2}\!+\!q\varphi_1 \Psi_{s1} \!+\!\mu_1\Psi_{s1}
\\
i{\hbar} {{\partial } \over {\partial t}} \Psi_{s2}\!&=&\!Te^{i {q \over {\hbar c}} \int^1_2 {\bf A} \cdot
d{\bf r}}\Psi_{s1}\!+\!q\varphi_2 \Psi_{s2}\!+\! \mu_2\Psi_{s2}.
\label{Kittel21}
\end{eqnarray}

Then, the current is given by 
\begin{eqnarray}
J\!=\! J_0
\sin \phi_s,
\end{eqnarray}
with $\phi_s$ given by
\begin{eqnarray}
\phi_s= \int_1^2 \left( {q \over {\hbar c}}  {\bf A} +{1 \over 2} \nabla \chi \right)\! \cdot \!d{\bf r}.
\label{phi}
\end{eqnarray}
By taking the same procedure as before, the AC Josephson frequency in Eq.~(\ref{phit2})  with $q=e$ is obtained. 

\begin{figure}
\begin{center}
\includegraphics[scale=0.5]{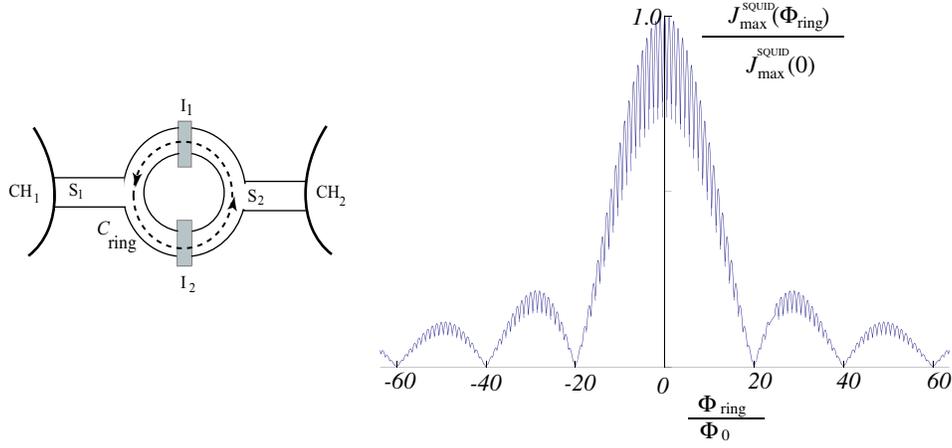}
\end{center}
\caption{Sketch of a SQUID structure and a plot of the maximum current through the SQUID structure $J^{\rm SQUID}_{\rm max}\!=\!{\rm Max} \{ J^{\rm even}_{\rm max},J^{\rm odd}_{\rm max} \} $ vs the magnetic flux through the ring $\Phi_{\rm ring}$. $\Phi_{\rm junc}$ is taken to be $\Phi_{\rm junc}=0.1 \Phi_{\rm ring}$.
}
\label{macroFraunAB}
\end{figure}

\section{SQUID oscillation in the spin-vortex superconuctivity}
We now show that the spin-vortex superconductivity explains the oscillation of the maximum current through a ring-shaped junction with respect to the external magnetic field.
Let us consider the SQUID structure with two identical SIS junctions shown in Fig.~\ref{macroFraunAB}. The total current is a sum of currents through the two junctions in the ring. If the magnetic flux through each junction area is neglected, the total current is calculated as
\begin{eqnarray}
J=J_0 \sin \phi_s +J_0\sin \left( \phi_s - 
{{\pi \Phi_{\rm ring}} /{ \Phi_0}} + w_{\rm ring}\pi \right),
\label{Jell}
\end{eqnarray}
where $\Phi_{\rm ring}$ is the flux through the ring region given by
\begin{eqnarray}
\Phi_{\rm ring}=\oint_{\rm C_{\rm ring}}  
{\bf A}\cdot d{\bf r},
\end{eqnarray}
and $w_{\rm ring}$ is an integer called the winding number of $\chi$ along ${\rm C_{\rm ring}}$ given by
\begin{eqnarray}
w_{\rm ring}={ 1 \over {2 \pi}}\oint_{\rm C_{\rm ring}}  
\nabla \chi \cdot d{\bf r}.
\end{eqnarray}

The current in Eq.~(\ref{Jell}) exhibits an interference pattern that depends on $\Phi_{\rm ring}$ and $w_{\rm ring}$. The phase difference is composed of the contribution from the Aharonov-Bohm phase from the magnetic filed $-\pi \Phi_{\rm ring}/\Phi_0$,
and that from the winding number of $\chi$ given by $w_{\rm ring}\pi$.
Note that in the Cooper-pair-tunneling case, the former is given by $-2\pi \Phi_{\rm ring}/\Phi_0$, and the latter is a multiple of $2\pi$ (i.e., $2\pi w_{\rm ring}$) \cite{Mercereau,Feynman}. 

The maximum current depends on whether $w_{\rm ring}$ is even or odd. By following the well-known derivations, it is calculated as
\begin{eqnarray}
J_{\max}^{\rm even}=J_{\max}^{\rm even}(0)\left| {{\sin \left(
{{\pi \Phi_{\rm junc}} \over {2\Phi_0}}\right)} 
\over {{\pi \Phi_{\rm junc}} \over {2\Phi_0}}} \right|
 \left| \cos \left(
{{\pi \Phi_{\rm ring}} \over {2\Phi_0}}\right) \right|,
\label{J1}
\end{eqnarray}
if $w_{\rm ring}$ is even;
and 
it is calculated as
\begin{eqnarray}
J_{\max}^{\rm odd}=J_{\max}^{\rm even}(0)\left| {{\sin \left(
{{\pi \Phi_{\rm junc}} \over {2\Phi_0}}\right)} 
\over {{\pi \Phi_{\rm junc}} \over {2\Phi_0}}} \right|
 \left| \sin \left(
{{\pi \Phi_{\rm ring}} \over {2\Phi_0}}\right) \right|,
\label{J2}
\end{eqnarray}
if $w_{\rm ring}$ is odd, 
where the effect of the flux through each junction region $\Phi_{\rm junc}$ is included.

If we take into account the change of $w_{\rm ring}$, the maximum current $J_{\max}^{\rm SQUID}$ is the largest of $J_{\max}^{\rm even}$ and $J_{\max}^{\rm odd}$. In Fig.~\ref{macroFraunAB}, a plot of $J_{\max}^{\rm SQUID}$ as a function of $\Phi_{\rm ring}$ is depicted. It exhibits a profile similar to the one observed in experiments \cite{Mercereau}, showing peaks separated by $\Phi_0$.

The corresponding maximum current for the $2e$-charge case is given by
\begin{eqnarray}
J_{\max}^{2e}=J_{\max}^{2e}(0) 
\left| {{\sin \left(
{{\pi \Phi_{\rm junc}} \over {\Phi_0}}\right)} 
\over {{\pi \Phi_{\rm junc}} \over {\Phi_0}}} \right|
 \left| \cos \left(
{{\pi \Phi_{\rm ring}} \over {\Phi_0}}\right) \right|;
\label{J3}
\end{eqnarray}
in this case, zero current is obtained at $\Phi_{\rm ring} / \Phi_0$=$1/2+n$, where $n$ is an integer in contrast to the $e$-charge case; therefore, the absence of these zero-points may be used as a support for the $e$-charge order parameter. Actually, the absence was observed in the experiment, but the reason for it has been attributed to a background current.

\begin{figure}
\begin{center}
\includegraphics[scale=0.5]{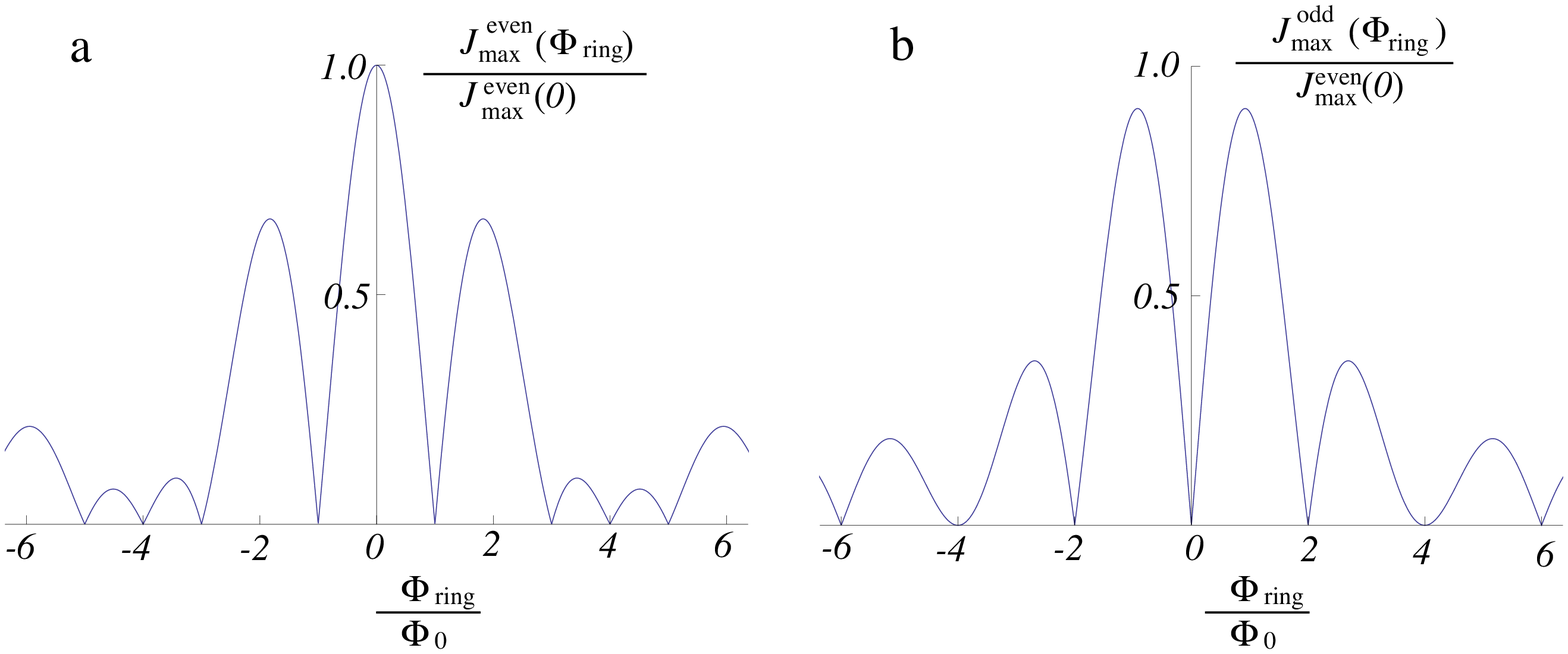}
\end{center}
\caption{a: Plot of $J_{\max}^{\rm even}$ as a function of $\Phi_{\rm ring}$. It shows a 
``$0$-SQUID" pattern. b: Plot of $J_{\max}^{\rm odd}$ as a function of $\Phi_{\rm ring}$; it shows a ``$\pi$-SQUID" pattern.
$\Phi_{\rm junc}$=$0.5 \Phi_{\rm ring}$ is used. If $\Phi_{\rm junc}$=$\Phi_{\rm ring}$ is used, the 
``$0$-SQUID" pattern becomes the edge-junction pattern and ``$\pi$-SQUID" pattern becomes the corner-junction pattern given in Ref.~\onlinecite{wollman}. }
\label{halfflux}
\end{figure}

It is worth noting that if the even-$w_{\rm ring}$ and odd-$w_{\rm ring}$ cases are separately realized, 
so-called, the ``$0$-SQUID"  and 
``$\pi$-SQUID" patterns seen in cuprates are obtained, respectively (see Fig.~\ref{halfflux}) \cite{wollman}.
The appearance of these patters are currently attributed to the d-wave symmetry of the order parameter; but it is explained due to the change of $w_{\rm ring}$ in the present theory.

\section{A possible relationship between the spin-vortex superconductivity and the BCS theory of superconductivity}
The present re-derivation of the AC Josephson frequency indicates that 
the BCS theory based on pairing electrons is incomplete as to the origin of the persistent current; persistent current carriers are electrons, not Cooper-pairs. 
Here, we try to relate the spin-vortex superconductivity to the BCS superconductivity by reinterpreting the phase of the order parameter; we assume that Cooper pairs are broken in the surface region and spin-vortices are formed. 

Let us consider the BCS state vector that for the situation where the phase $\theta$ varies over the sample as
\begin{eqnarray}
|\Psi_{\rm BCS} \rangle=\prod_{{\rm a}, {\bf k}}\left( u_{{\rm a} {\bf k}}+e^{i\theta_{\rm a}}v_{{\rm a} {\bf k}}c^{\dagger}_{{\rm a} {\bf k} \uparrow}c^{\dagger}_{{\rm a} -{\bf k} \downarrow} \right)|{\rm vac}\rangle.
\label{BCS1}
\end{eqnarray}
Here, the whole system is devided into small cells, and each cell (indexed by a) is assumed to be described by the BCS-type state vector given in Eq.~(\ref{BCS0}) \cite{BCS}. The real parameters $u_{ {\rm a} {\bf k}}$ and $v_{{\rm a} {\bf k}}$ satisfy $u^2_{{\rm a} {\bf k}}+v^2_{{\rm a} {\bf k}}=1$.

In the conventional theory, $\theta$ is thought to arise from the fluctuation of the number of Cooper pairs. The equation to obtain $\theta$ and $\rho$ is the Ginzburg-Landau equation\cite{Ginzburg} that are derived from the BCS theory \cite{Gorkov}; however, this derivation yields charge $2e$ for the carrier, thus, contradicts the observation that the charge on the carrier is $e$.
In the following, we derive the equations of motion for $\theta$ and $\rho$ in a different maner, and relate $\theta$ to $\chi$ that arises from the spin-vortex formation.

We employ the following Lagrangian
\begin{eqnarray}
{\cal L}_{\rm BCS}&=&\langle  \Psi_{\rm BCS} | i\hbar {{\partial } \over {\partial t}}-H_{\rm BCS}-e\sum_{{\rm a},  {\bf k}, \sigma }\varphi_{\rm a} c^{\dagger}_{{\rm a} {\bf k} \sigma}c_{{\rm a} {\bf k} \sigma}| \Psi_{\rm BCS} \rangle
\nonumber
\\
&=&-\hbar \sum_{\rm a} \rho_{\rm a}^{\rm pair} \dot{\theta}_{\rm a}
-{\cal H}_{\rm BCS} -2e \sum_{\rm a} \rho_{\rm a}^{\rm pair} \varphi_{\rm a},
\end{eqnarray}
 where $\rho_{\rm a}^{\rm pair}=\sum_{\bf k} u_{{\rm a} {\bf k}}^2$ is the number of Cooper pairs in the cell a, and ${\cal H}_{\rm BCS}=\langle  \Psi_{\rm BCS} |H_{\rm BCS} | \Psi_{\rm BCS} \rangle$ is the energy functional for the BCS Hamiltonian.
 From this Lagrangain the following equations of motion are obtained;
\begin{eqnarray}
\dot{\rho}^{\rm pair}&=&{ {1} \over {\hbar}} {  {\delta {\cal H}_{\rm BCS}} \over {\delta \theta}}, 
\label{EOMBCS1}
\\
\dot{\theta}+{{2e} \over \hbar} \varphi &=&-{1 \over \hbar} {{\delta {\cal H_{\rm BCS}}} \over {\delta \rho^{\rm pair}}}.
\label{EOMBCS2}
\end{eqnarray}

We assume that even some of Cooper pairs are broken and spin-vortices are created, we may still use the BCS-type state vector as an approximate one if the phase $\chi$ is included. This means that an approximate state vector is written as
\begin{eqnarray}
|\Psi_{\rm BCS} [\chi]\rangle=\prod_{{\rm a}, {\bf k}}\left( u_{{\rm a} {\bf k}}+v_{{\rm a} {\bf k}}a^{\dagger}_{{\rm a} {\bf k}}b^{\dagger}_{{\rm a} {\bf k} } \right)|{\rm vac}\rangle, 
\label{BCS2}
\end{eqnarray}
where new operators $a_{{\rm a} {\bf k}}$ and 
$b_{{\rm a} {\bf k}}$ are defined by
\begin{eqnarray}
\left( \begin{array}{c}
a_{{\rm a} {\bf k}}\\
b_{{\rm a} {\bf k}}
\end{array}
\right)={ {e^{ i{\chi_{\rm a} \over 2}}} \over \sqrt{2} } \left(
\begin{array}{cc}
e^{i {\xi_{\rm a} \over 2}} & e^{-i {\xi_{\rm a} \over 2}} \\
-e^{i {\xi_{\rm a} \over 2}} & e^{-i {\xi_{\rm a} \over 2}} 
 \end{array}
 \right) \left( \begin{array}{c}
c_{{\rm a} {\bf k} \uparrow}\\
c_{{\rm a} -{\bf k} \downarrow}
\end{array}
\right),
\end{eqnarray}
which corresponds to the transformation in Eq.~(\ref{transform1}).

The above state vector is actually rewritten as
\begin{eqnarray}
|\Psi_{\rm BCS}[\chi] \rangle=\prod_{{\rm a}, {\bf k}}\left( u_{{\rm a} {\bf k}}+e^{-i\chi_{\rm a}}v_{{\rm a} {\bf k}}c^{\dagger}_{{\rm a} {\bf k} \uparrow}c^{\dagger}_{{\rm a} -{\bf k} \downarrow} \right)|{\rm vac}\rangle.
\end{eqnarray}
The comparison with Eq.~(\ref{BCS1}) gives $\theta=-\chi$. By employing this reinterpretation many practical problems may be handled. However, the true equations of motion are given in Eqs.~(\ref{EOM0}) and (\ref{EOM1}), not in Eqs.~(\ref{EOMBCS1}) and (\ref{EOMBCS2}). 

It is known that the transition temperature $T_{\rm c}$ calculated as the temperature for the Cooper-pair formation explains T$_{\rm c}$ in many superconducting materials. However, the present work indicates that the Cooper-pair is not the persistent current carrier. In this respect, it may be worth noting that a study of correlations between normal-state properties and the occurrence of superconductivity indicates that properties assumed to be important within BCS theory rank lowest in predictive power
regarding whether a material is or is not a superconductor.\cite{Hirsch}

The above facts may be explained if we assume that fragile germs of superconductivity already exist above T$_{\rm c}$, and whether they exist or not is related to the occurrence of superconductivity. In this case, the transition temperature is the temperature where the germs become stable and coherently connected over the sample, that presumably occurs when the energy gap is formed. In this respect it is worth mentioning that the germs of superconductivity (or spin-vortices) are speculated to exist in the pseudogap phase above T$_{\rm c}$ in cuprates \cite{koizumi2}. The spin-vortices may be stable only in the surface region in the conventional superconductors, but even in the bulk in the cuprates due to the two-dimensionality of the conduction CuO$_2$ plane. The anomalous pseudogap phase and the high T$_{\rm c}$ value in the cuprates may be related to the existence of rather stable spin-vortices in the bulk \cite{koizumi2}. 

\section{concluding remarks}
We have shown that the AC Josephson frequency $2|e|V/h$ actually means that the charge on the superconducting order parameter is $e$. The conventional  interpretation that it is the evidence of the tunneling of $2e$-carriers is reversed. 
The present result indicates that although the BCS theory based on the Cooper-pair formation explains bulk properties of superconductors very well, it fails to explain the persistent current flow which occurs predominantly in the surface region. We speculate that some of the Cooper-pairs are broken or Cooper-pairs are not formed in the surface region, and spin-vortices are created there; then, the phase of the order parameter arises as a Berry phase effect due to the spin-vortex formation. This also indicates that Cooper pairs are not an absolutely necessary ingredient of superconductivity \cite{koizumi2}.

We would like to mention that there are some experimental and theoretical indications that spin-vortices exist in high-transition temperature cuprate superconductors \cite{koizumi2}. The high T$_{\rm c}$ observed in them may be related to the fact that the two-dimensionality of the current flowing CuO$_2$ plane is advantageous for the stabilization of spin-vortces. 

In the superfluidity of Bose particles, equations of motion in Eqs.~(\ref{boseA}) and (\ref{bose}) are used. In this case, however, the phase of the order parameter cannot be attributed to spin-vortices since the spin degrees-of-freedom is absent. Nevertheless, such equations are obtained by using two components that arise in this system due to the Bose-Einsein condensation.
The one of the two components is the condensate state and the other is the non-condensate states; the phase of the order parameter may be attributed to the relative phase between them. We will briefly describe this possibility below.

It is known that the ground state vector for the dilute Bose gas is given by
\begin{eqnarray}
|\Psi_{N}\rangle={\rm const.} \left( \alpha_0^{\dagger}\alpha_0^{\dagger}- \sum_{\bf k} c_{\bf k} \alpha_{\bf k}^{\dagger}\alpha_{-{\bf k}}^{\dagger}\right)^{N/2} |{\rm vac} \rangle,
\end{eqnarray}
where $ \alpha_0^{\dagger}$ is the creation operator for the condensate state, and
$\alpha_{\bf k}^{\dagger}$ describes that for other states; $c_{\rm k}$ are parameters and $N$ is the total number of the particles \cite{Leggett1}.

We divide the whole system into small cells, and each cell (indexed by a) is assumed to be described by the above state vector. When the phase of the condensate wave function varies relative to the rest, slowly, over the cells, we have  
\begin{eqnarray}
|\Psi_{N}[\theta] \rangle={\rm const.} \prod_{\rm a}\left( \alpha_{{\rm a} 0}^{\dagger}\alpha_{{\rm a} 0}^{\dagger}e^{i 2 \theta_{\rm a}}- \sum_{\bf k} c_{{\rm a} {\bf k}} \alpha_{{\rm a} {\bf k}}^{\dagger}\alpha_{{\rm a} -{\bf k}}^{\dagger}\right)^{N_{\rm a}/2} |{\rm vac} \rangle,
\label{bosevelo}
\end{eqnarray}
where $\theta_{\rm a}$ and $N_{\rm a}$ are the relative phase and number of particles in the cell a.

Let us employ the following Lagrangian, 
\begin{eqnarray}
{\cal L}_B=\langle \Psi_{N}[\theta]| \left(i \hbar { {\partial} \over {\partial t} } - H_B \right)|\Psi_{N}[\theta] \rangle,
\label{lagbose}
\end{eqnarray}
where $H_B$ is the Hamiltonian for the Bose system. We denote the energy functional of the system as
\begin{eqnarray}
{\cal H}_B=\langle \Psi_{N}[\theta]| H_B |\Psi_{N}[\theta] \rangle,
\label{hambose}
\end{eqnarray}
and the number of particles in the condensate state in the cell a  as
\begin{eqnarray}
\rho_{\rm a}=\langle \Psi_{N}[\theta]|\alpha_{{\rm a} 0}^{\dagger}\alpha_{{\rm a} 0}
|\Psi_{N}[\theta] \rangle.
\end{eqnarray}

From the Lagrangian in Eq.~(\ref{lagbose}), the following
equations of motion are obtained;
\begin{eqnarray}
\dot{\rho}&=&{ {1} \over {\hbar}} {  {\delta {\cal H}_B} \over {\delta \theta}}=-{ {1} \over {\hbar}} \nabla \cdot {  {\delta {\cal H}_B} \over {\delta \nabla \theta}}, 
\label{bose2A}
\\
\dot{\theta}&=&-{1 \over \hbar} {{\delta {\cal H}_B} \over {\delta \rho}}.
\label{bose2}
\end{eqnarray}

The equations of motion in Eqs.~(\ref{bose2A}) and (\ref{bose2}) correspond to those given in Eqs.~(\ref{boseA}) and (\ref{bose}). The equation (\ref{bose2A}) indicates that superfluid velocity is proportional to $\nabla \theta$; if we neglect the
non-condensate component of the state vector in Eq.~(\ref{bosevelo}), the phase $\theta$ is the phase of the condensate wave function. Then, the present superfluid velocity agrees with that is given by A. J. Leggett \cite{Leggett2}. In the present view, the two superfluid phenomena are unified in the common mathematical structure where
the particle states are locally given in two components; spin degrees-of-freedom for the superconductivity case, and condensate and non-condenstate states in the Bose-Einstein condensation case. This view is markedly different from the view that both are due to the $U(1)$ gauge symmetry breaking \cite{Anderson66}.

In conclusion, we have derived the AC Josephson frequency in a gauge invariant manner with including a battery contact effect. The result indicates that the persistent current is a flow of electrons, thus, reverses the current understanding of the origin of the phase of the order-parameter. We have presented an alternative origin of the phase of the order parameter; we have argued that the Berry phase that arises from spin-vortices is the origin. 

The author thanks S. Sugano, T. Sasada, Y. Takada, A. Fujimori, H. Aoki, N. Nagaosa, T. Yanagisawa, H. Shiba, K. Ueda, M. Oshikawa, T. Kato for helpful comments.


\end{document}